\documentstyle[12pt,aasms4]{article}

\begin{document}

\title{Gamma-Ray Burst Arrival-Time Localizations: Simultaneous Observations by
Ulysses, Pioneer Venus Orbiter, SIGMA, WATCH, and PHEBUS}

\author{K. Hurley}
\affil{Space Sciences Laboratory, University of California, Berkeley, CA 94720-7450}
\authoremail{khurley@sunspot.ssl.berkeley.edu}

\author{J. Laros}
\affil{Lunar and Planetary Laboratory, University of Arizona, Tucson, AZ 85715}

\author{S. Brandt, E. E. Fenimore, R. W. Klebesadel, J. Terrell}
\affil{Los Alamos National Laboratory, Los Alamos, NM 87545}

\author{T. Cline}
\affil{NASA Goddard Space Flight Center, Code 661, Greenbelt, MD 20771}

\author{C. Barat, M. Boer, J.-P. Dezalay}
\affil{CESR, F-31029 Toulouse Cedex, France}

\author{R. Sunyaev, O. Terekhov, A. Kuznetsov, S. Sazonov}
\affil{IKI, 117810 Moscow, Russia}

\author{N. Lund}
\affil{Danish Space Research Institute, DK-2100 Copenhagen, Denmark}

\author{A. Claret, J. Paul}
\affil{CEA Saclay, Orme des Merisiers, F-91191 Gif-sur-Yvette Cedex, France}

\author{A. Castro-Tirado\altaffilmark{1}}
\affil{Laboratorio de Astrof\'{\i}sica Espacial y F\'{\i}sica 
Fundamenta (LAEFF-INTA), P.O. Box 50727, E-28080 Madrid, Spain}

\altaffiltext{1}{Instituto de Astrof\'{\i}sica de Andaluc\'{\i}a (IAA-CSIC),
P.O. Box 03004, E-18080 Granada, Spain}

\begin{abstract}

Between the launch of the Ulysses spacecraft in 1990 October and the entry of
Pioneer Venus Orbiter (PVO) into the atmosphere of Venus in 1992 October, concurrent
coverage by Ulysses, PVO, the WATCH experiments aboard the Granat and EURECA spacecraft,
and the SIGMA and PHEBUS experiments aboard the Granat spacecraft was obtained for
numerous gamma-ray bursts.  15 of them were detected by 3 or more instruments 
on spacecraft separated by distances of several AU, and could therefore be accurately localized by triangulation.
In some cases independent, accurate locations were obtained by SIGMA and/or
WATCH.  We present these localizations, which range in area from 0.9 to 530 
arcminutes$^2$.

\end{abstract}


\section{Introduction}

A knowledge of the precise locations of cosmic gamma-ray bursts (GRBs) is important
for many studies.  When obtained rapidly, they allow multi-wavelength counterpart
searches to be carried out, which have led to the discovery of fading radio and optical
counterparts.  After days to months, these fading counterparts are unlikely to be
detected, but precise locations, even obtained years later, are useful for statistical
studies, such as clustering, searches of cataloged objects for possible associations,
and host galaxy limits.  A number of GRBs have had their redshifts spectroscopically measured or
constrained (e.g., Metzger et al. 1997; Kulkarni et al. 1998; Djorgovski et al. 1998), and the results 
establish a cosmological origin for them, i.e. z$\gtrsim$1.  More recently, however, one
burst has been shown to be spatially and temporally coincident with a nearby supernova
(Galama et al. 1998), indicating that GRBs may be a diverse phenomenon, and that
their counterparts may not all be faint galaxies which would be difficult to discern
at late times in a relatively large error box.  This paper is the fifth in a series presenting GRB
localizations by triangulation between spacecraft in the 3rd Interplanetary Network (IPN),
which are separated by several thousand light-seconds.  
The first two presented results obtained with the \it Ulysses \rm, Compton Gamma-Ray
Observatory, and Pioneer Venus Orbiter (PVO) or Mars Observer spacecraft (Laros et al. 1997,
1998).
The third and the fourth are the \it Ulysses \rm supplements to the BATSE 3B and 4Br
catalogs (Hurley et al. 1999a,b).  This paper presents results obtained with the \it Ulysses \rm, PVO, Granat,
and EURECA spacecraft.  The Granat experiments involved were WATCH, PHEBUS, and/or SIGMA.
The EURECA experiment was WATCH.

\section{Instrumentation}

All of the instrumentation used to obtain the data presented here is based on
scintillation crystals, and all the instruments have been described in detail
elsewhere.  We review each briefly.

The \it Ulysses \rm GRB detector (Hurley et al. 1992) consists of two 3 mm thick
hemispherical CsI scintillators
with a projected area of about 20 cm$^2$ in any direction.  Its nominal
energy range is 25-150 keV.  GRB time histories are recorded with time resolutions
which range from about 8 ms (in a triggered mode) to 0.5 - 2 s (in real time
modes).  The detector is mounted
on a magnetometer boom far from the body of the spacecraft, and has a practically
unobstructed view of the full sky.  The \it Ulysses \rm orbit is heliocentric,
with a 5 AU aphelion.  The instrument has no inherent burst localization
capability.

PVO had two burst detectors, consisting of 3.8 cm diameter by 3.2 cm long CsI scintillators,
operating in the 100-2000 keV energy range.  Time histories were recorded
with resolutions ranging from 1/4096 s in time-to-spill mode, to 12/1024 s in 
triggered mode, to 16 s in real
time mode.  The spacecraft was in orbit around
the planet Venus for the observations reported here.  Like the \it Ulysses \rm
detector, it had no inherent directional capability.  Further details may be
found in Klebesadel et al. (1980).

The SIGMA telescope was a coded mask imaging system capable of localizing sources to
arcminute accuracy within the fully coded field of view.  However, the bursts 
described in the present paper were observed in the sidelobes, and the images
were partially coded, leading to accuracies in the 10's of arcminutes range and
above.  The localization
procedure is described in Claret et al. (1994).
GRB time histories were generally recorded by the SIGMA anticoincidence
system, which operated in the energy range above several hundred keV (the precise
threshold for any given photon interaction depends on the location of the
interaction in the shield).  The time resolution was variable, depending on
the count rate (time-to-spill mode) but typically was around 100 ms and greater.
  SIGMA was mounted on
the Granat spacecraft, which was in a highly eccentric Earth orbit with apogee $>$ 70000
km.

The WATCH instrument was also aboard the Granat spacecraft.  Based on
a novel rotating modulation collimator technique, the WATCH detectors surveyed 
~80\% of the sky, and localized bursts to elliptical error boxes, which may
be approximated by circles whose 3 $\sigma$ radii are 0.2 - 1.6 $\arcdeg$.  
The localization accuracy depends, among other things, on the accuracy with
which the attitude of the Granat spacecraft can be reconstructed.  In general,
the spacecraft attitude was derived from the star tracker, which was part
of the SIGMA instrument, and when it was operating the uncertainties were negligible as far as
the burst locations in this catalog are concerned.  However, WATCH detected
some bursts at times when SIGMA was off, and only the predicted spacecraft
attitude is known.  The spacecraft actually oscillates slowly about this
predicted position, with peak-to-peak amplitudes of 30 - 40 $\arcmin$, 
independently about three axes.  In these cases, the attitude was reconstructed
by fitting the positions of bright X-ray sources in the WATCH data for periods
approximately 30 m long about the time of the burst.  This procedure recovers
the secular drift associated with solar motion, but not the oscillations, and
a systematic uncertainty of 0.5 $\arcdeg$ was assumed to account for them.    The energy
range was 8 - 150 keV and the time resolution ranged up to approximately 0.8
s.  WATCH/Granat is described in Sazonov et al. (1998).  A
similar instrument was also launched later aboard the EURECA spacecraft 
into low Earth orbit (Lund, 1985).  

Finally, the PHEBUS experiment was also included in the Granat payload (Barat et al. 1988; Terekhov et al. 1991), consisting
of six 12 cm. long by 7.8 cm. diameter BGO detectors oriented along the axes of a Cartesian
coordinate system, and operating in the 100 keV - 100 MeV energy range, with 1/128 s to
1/32 s time resolution.  By comparing
the count rates of the various detectors, it is possible to obtain an approximate
source location; the accuracies vary depending on the burst, but are in the several
10's of degrees range and above.  Although quite coarse, this information proved to
be very valuable for some of the bursts described here (see below).  The spacecraft
attitude uncertainties discussed above are negligible compared to the PHEBUS
localization uncertainties.

At the time the bursts in this catalog were detected, the interest in providing small error boxes
rapidly was recognized.  However, the mission designs, in some cases already 15 years old,
did not always allow for this.  Nevertheless, in three cases (GRB910219, GRB911016, and
GRB920714) localizations were done rapidly enough to allow imaging of the fields
down to 18th magnitude within three days, although no optical counterparts were found (Castro-Tirado et al.
1994).

\section{GRB Localization}

The precise error boxes presented here have been derived by triangulation, or
arrival-time analysis between widely separated spacecraft.  (``Widely separated''
here means distances of several AU.)  This method consists
of analyzing the time histories of a GRB as recorded by two spacecraft in order
to determine the most likely time difference and its statistical uncertainty.
This analysis is done using a $\chi^2$ statistic (e.g. Hurley et al. 1999a; Laros et al. 1997).  There is, however, an important difference
between the events presented here and those presented in previous catalogs, which
requires further explanation.

GRB time histories are energy-dependent.  A time history taken in the 25-100 keV
\it Ulysses \rm energy range may differ from that taken in the PVO 100-2000 keV range.
The magnitude of this difference varies considerably from event to event, and
can easily be judged in, say, the $\chi^2$ technique (Hurley et al. 1999a), where
the goodness-of-fit is reflected in the value of $\chi^2$ per degree of freedom.
When the ``fit'' between two time histories is poor, the estimate of the statistical uncertainty
in the time difference may become unreliable.  This, in turn, renders the annulus
width estimates, and hence the confidence value for the error box, suspect.
In previous GRB location catalogs, we have been able to avoid this problem
by comparing time histories in the same, or very similar energy ranges.  Thus
in the \it Ulysses \rm /BATSE catalogs, 25-150 keV Ulysses time histories were
compared to 25-100 keV BATSE time histories, and the fits were generally satisfactory
(Hurley et al. 1999a,b).  In the \it Ulysses \rm /BATSE/PVO catalog, the 25-150 \it
Ulysses \rm time histories were again compared to the 25-100 keV BATSE time histories,
but the 100-2000 keV PVO time histories were compared to the $>$100 keV BATSE time histories.
 
In the present catalog, we have four instruments - \it Ulysses \rm, PVO, PHEBUS, and SIGMA - 
which recorded their time histories in a single energy range which was different from all
the others.  Some of these energy ranges, e.g. \it Ulysses' \rm and SIGMA's, do not
even overlap.  We have therefore used the following techniques to assure that the
error boxes are conservatively estimated.  First, we have used the PHEBUS 100 keV -
100 MeV time history for the comparisons
instead of the SIGMA one (every SIGMA event in this catalog was also observed by PHEBUS).  
Second, we perform the following internal consistency check.

Let $\rm \delta T_{i-j}$ be the difference in arrival times between spacecraft
i and spacecraft j.  Let the subscript c denote the calculated values, and t the true
values, unknown to the experimenter.    Then for a network of three spacecraft,
$\rm \delta T_{1-2,t} + \delta T_{1-3,t} + \delta T_{3-2,t} \equiv 0$.  In general, the
sum of the calculated values will not be zero, due to a combination of
statistical and systematic errors:
$\rm \Delta \equiv \delta T_{1-2,c} + \delta T_{1-3,c} + \delta T_{3-2,c} \neq 0$.  Let
$\rm \sigma(\delta T_{i-j}$) be the statistical error associated with $\rm \delta T_{i-j,c}$.
(We have shown in Hurley et al., 1999a, that the error distribution should be approximately
normal).   In those cases where $\Delta$ is incompatible with the values of 
$\rm \sigma(\delta T_{i-j}$), we increase them appropriately.

Finally, we note that the events in this catalog were all observed by just three
widely separated spacecraft.  Triangulation therefore yields two possible intersection
points for the annuli.  We have generally used the localization capabilities of WATCH and SIGMA
to identify the correct intersection.  In those cases where no WATCH or SIGMA data was
available, we have used the PHEBUS location capability to identify the intersection.

\section{GRB Locations}

Table 1 gives the dates and times of the bursts and identifies the spacecraft which observed
them and their operating modes.  In some cases, bursts were observed by additional near-Earth spacecraft, such
as DMSP (The U.S. Air Force Defense Meteorological Satellite Program: Terrell et al. 1998).  
These data were consistent with those of other near-Earth spacecraft, however, and their use    did not constrain the error boxes further.  Also,
some events were localized to two alternate error boxes which could not be
distinguished using the directional response of any of the instruments.  These
are not discussed further\footnote{Full details of these events in particular,
and all bursts in general localized by the IPN, 
may be found at http://ssl.berkeley.edu/ipn3/index.html.}.  This table also indicates
which bursts were observed when the SIGMA star tracker was off and the spacecraft
attitude could not be precisely determined, as discussed above.

For the bursts in Table 1, Table 2 gives the corners of the error box, the center
of the error box, its area, and its maximum dimension.  The epoch for the
coordinates is J2000.  In general, the smallest possible error box derived 
from triangulation using 3 spacecraft will be defined by 
4 or 6 corners from the intersection of 3 annuli (depending on the width of
the annuli), but in some cases, as noted below in
Table 2 and the figure captions, grazing
intersections may reduce this number.  The coordinates have been corrected to the
heliocentric frame (the equivalent of the aberration correction - see Hurley et al. 1999a),
and supercede all previous data on these bursts.  Some of the error boxes are shown
in figures 1 through 7.  In two cases the WATCH and IPN annuli are only
marginally compatible (figures 4 and 6); it is thought that the cause
is 1) the imprecisely known
Granat spacecraft attitude, which may result in a systematic underestimate of the total
WATCH error circle radius, and/or 2) the approximation of the elliptical WATCH locations
by circles.

\section{Conclusions}

IPN error box areas are comparable in size to, or in some cases much smaller than 
those that can be derived rapidly from wide field X-ray cameras such as the one
on board BeppoSAX ($\approx 10 \arcmin$ error circle radius - e.g. Costa et al 1997).  
For most of the events in this catalog, the initial error boxes were circulated to the
astronomical community with delays which were considerably greater than those
that can be achieved by BeppoSAX.  However, the fact that fading optical transients
can be detected in the BeppoSAX wide field camera error circles even several days
after the burst means that an IPN which can deliver small error boxes on $\approx$
day timescales will be useful for counterpart identifications.  Such a network now
exists, consisting of \it Ulysses, \rm BATSE, and the Near Earth Asteroid Rendezvous
mission (Cline et al. 1999).

Finally, we note that over 50 bursts were detected by \it Ulysses, \rm WATCH, and
in some cases other spacecraft, which will provide error boxes with areas of
several hundred arcminutes$^2$.  Publication of these events is in preparation. 

\acknowledgments

KH is grateful to JPL Contract 958056 for Ulysses operations.

\clearpage

\figcaption{The IPN annuli and the WATCH 3 $\sigma$ error circle for GRB 901204.\label{Fig. 1}}

\figcaption{The IPN annuli, the WATCH 3$\sigma$ error circle, and the SIGMA 1, 2,
and 3 $\sigma$ error contours for GRB910122.\label{fig2}}

\figcaption{The IPN annuli and the WATCH 3 $\sigma$ error circle for GRB910219.\label{fig3}}

\figcaption{The IPN annuli and the WATCH 3 $\sigma$ error circle for GRB910310.  The
major contribution to the WATCH error circle uncertainty in this case is the
poorly known Granat spacecraft attitude.  This systematic error may be underestimated
here.\label{fig4}}

\figcaption{The BATSE 1 $\sigma$ error circle (left, Meegan et al. 1996), the WATCH 3 $\sigma$ error circle (right),
and the IPN annuli for GRB911016.  The narrow annulus is fully contained within
the wider one; its intersection with the WATCH error circle defines the error box.\label{fig5}}

\figcaption{The BATSE 1 $\sigma$ error circle (bottom right, Meegan et al. 1996), the WATCH 3 $\sigma$
error circle (top left), the IPN annuli (intersecting at grazing incidence), 
and the SIGMA 1 and 2 $\sigma$ error
contours for GRB920714.  \label{fig6}}

\figcaption{The WATCH 3 $\sigma$ error circle, the SIGMA 1, 2, and 3 $\sigma$
error contours, and the IPN annuli for GRB920723.\label{fig7}}

\clearpage
\small
\begin{deluxetable}{cccccccc}
\tablecaption{Bursts in this catalog}
\tablehead{
\colhead{Date}&\colhead{ECT\tablenotemark{a}} & \colhead{\it Ulysses}
& \colhead{BATSE} & \colhead{PVO} & \colhead{WATCH} & \colhead{SIGMA} & 
\colhead{PHEBUS} 
}
\startdata

 4 DEC 90\tablenotemark{b} & 09:42:52 &  YES\tablenotemark{c} & N/O\tablenotemark{d} & YES & YES & N/O & YES   \nl 
 6 JAN 91\tablenotemark{e} & 16:39:57 &  YES & N/O & RI\tablenotemark{f}  & NO\tablenotemark{g}  & NO & YES   \nl 
22 JAN 91 & 15:13:49 &  YES & N/O & YES & YES & YES & YES   \nl
11 FEB 91 & 03:25:22 &  YES & N/O & RI  & NO  & NO & YES   \nl
19 FEB 91\tablenotemark{e} & 11:45:24 &  YES & N/O & YES & YES & NO & N/O   \nl
10 MAR 91\tablenotemark{b} & 13:02:05 &  YES & N/O & YES & YES & N/O & N/O   \nl
 2 APR 91 & 14:28:15 &  RI  & N/O & RI  & NO  & YES & YES   \nl                                                                         
17 APR 91\tablenotemark{e} & 20:07:32 &  YES & N/O & YES & NO  & YES & YES \nl
17 MAY 91\tablenotemark{b} & 05:02:43 &   RI & NO  & YES & YES\tablenotemark{h} & YES & YES  \nl
16 OCT 91\tablenotemark{b} & 11:01:36 &  RI  & YES & RI  & YES & N/O & YES  \nl
18 OCT 91 & 05:32:15 &  YES & NO  & YES & NO  & NO & YES  \nl
22 DEC 91 & 15:00:10 &  RI  & NO  & RI  & NO  & NO & YES  \nl
19 MAY 92 & 16:31:53 &  RI  & YES & YES & NO  & N/O & YES  \nl
14 JUL 92 & 13:04:29 &  RI  & YES & RI  & YES & YES & YES  \nl                                                      
\tablebreak
23 JUL 92 & 20:03:08 &  YES & NO  & YES & YES & YES & YES  \nl
 4 OCT 92 & 14:00:21 &  YES & NO  & YES & YES & NO & N/O  \nl

\enddata
\tablenotetext{a}{Earth crossing time, UT}
\tablenotetext{b} {Granat attitude missing}
\tablenotetext{c}{Burst was observed in a triggered (high time resolution) mode}
\tablenotetext{d}{Burst was not observable due to, e.g. a data gap, spacecraft not yet launched, etc.}
\tablenotetext{e}{Two possible locations}
\tablenotetext{f}{Burst was observed in untriggered mode, as a rate increase (low time resolution)}
\tablenotetext{g}{Data were available, clean, and complete, but burst was not observed}
\tablenotetext{h}{Burst was observed by WATCH, but could not be localized}

\end{deluxetable}

\normalsize

\clearpage
\begin{deluxetable}{ccccccc}
\tablecaption{Error boxes of the bursts in Table 1} 
\tablehead{
\colhead{} & \multicolumn{2}{c}{Error box corners} & 
\multicolumn{2}{c}{Error box center} & \colhead{Error box} & \colhead{Maximum error} \\
\colhead{Date} & \colhead{$\alpha(2000)$} & \colhead{$\delta(2000)$} & 
\colhead{$\alpha(2000)$} & \colhead{$\delta(2000)$}  & \colhead{area,} 
& \colhead{box dimension,} \\
\colhead{} & \colhead{} & \colhead{} & \colhead{} & \colhead{} & \colhead{arcmin.$^2$} & \colhead{arcmin.}   \\
}
\startdata
901204\tablenotemark{a} &
     296.197 &      37.747 & 296.483 & 37.586 & 43 & 35 \\
  &   296.766 &      37.423 & & &  \\
  &   296.363 &      37.668 & & & \\
  &   296.602 &       37.503 & & & \\
  &  296.173 &       37.749 & & & \\
  &  296.791 &     37.421 & & & \\

910122\tablenotemark{a} & 296.918 & -70.681 & 296.756 & -70.646 & 18 & 10 \\
       & 296.595 & -70.612 &            &            &    &    \\
       & 296.674 & -70.660 &            &            &    &    \\
       & 296.838 & -70.633 &            &            &    &    \\
       & 297.000 & -70.667 &            &            &    &    \\
       & 296.512 & -70.626 &            &            &    &    \\

910219\tablenotemark{b} & 
      213.731 & 58.671 & 213.694 & 58.688 & 7.3 & 4.7 \\
 &    213.657 & 58.705 &            &           &     &     \\
 &    213.723 & 58.710 &            &           &     &     \\
 &    213.665 & 58.666 &            &           &     &     \\
 &    213.701 & 58.649 &            &           &     &     \\
 &    213.687 & 58.727 &            &           &     &     \\

\tablebreak

910310\tablenotemark{b} & 
       184.358 &  7.266  & 184.304 & 7.196 & 63 & 38 \\
 &     184.249 & 7.125 &            &          &    &    \\
 &    184.198 &  6.921 &            &          &    &    \\
 &    184.405 &  7.462 &            &          &    &    \\
 &    184.424 &  7.480 &            &          &    &    \\
 &    184.178 &  6.901 &            &          &    &    \\

910402\tablenotemark{a} & 
         77.612 & 13.611 & 77.629 & 13.690 & 35 & 14 \\
       & 77.647 & 13.768 &           &           &    &    \\
       & 77.598 & 13.675 &           &           &    &    \\
       & 77.661 & 13.704 &           &           &    &    \\
       & 77.631 & 13.571 &           &           &    &    \\
       & 77.627 & 13.810 &           &           &    &    \\

910517\tablenotemark{a} & 
         150.475 & -42.876 & 150.602 & -42.780 & 236 & 92 \\
       & 150.730 & -42.693 &            &            &     &    \\
       & 149.659 & -43.107 &            &            &     &    \\
       & 151.546 & -42.447 &            &            &     &    \\
       & 151.545 & -42.447 &            &            &     &    \\
       & 149.659 & -43.107 &            &            &     &    \\

911016\tablenotemark{c} & 297.996 & -5.386 & 298.137  & -4.811    & 530 & 70   \\         
       & 298.151 & -4.220 &          &           &     &      \\
       & 298.148 & -5.205 &          &           &     &      \\
       & 298.251 & -4.434 &          &           &     &      \\

\tablebreak

911018\tablenotemark{a} & 
         5.468 & 31.957 & 6.009    & 31.658    & 50 & 74 \\
       & 6.542 & 31.353 &          &           &    &    \\
       & 5.401 & 31.992 &          &           &    &    \\
       & 6.608 & 31.316 &          &           &    &    \\
       & 6.486 & 31.397 &          &           &    &    \\
       & 5.526 & 31.914 &          &           &    &    \\

911222\tablenotemark{a} &
         87.139 & 13.137 & 87.084 & 14.062 & 200 & 118 \\
       & 87.005 & 14.938 &        &        &     &     \\
       & 87.159 & 13.274 &        &        &     &     \\
       & 87.023 & 15.111 &        &        &     &     \\
 
920519\tablenotemark{a} & 
         321.422 &  44.221 & 321.254 & 44.137  & 21  & 23 \\
       & 321.485 &  44.228 &            &            &     &    \\
       & 321.024 &  44.046 &            &            &     &    \\
       & 320.087 &  44.053 &            &            &     &    \\

920714\tablenotemark{d} & 
         220.826    & -30.721    &   220.857  &  -30.610   & 36 & 20 \\
       & 220.897    & -30.506    &            &            &    &    \\   
       & 220.848    & -30.607    &            &            &    &    \\

\tablebreak

920723\tablenotemark{a} &
         287.128 & 27.216  & 287.142 & 27.232  & 0.9 & 3.2 \\
       & 287.155 & 27.248  &            &            &     &     \\
       & 287.126 & 27.210  &            &            &     &     \\
       & 287.157 & 27.255  &            &            &     &     \\
       & 287.148 & 27.249  &            &            &     &     \\
       & 287.135 & 27.215  &            &            &     &     \\

921004\tablenotemark{e} & 
         219.244 & 34.180 & 219.236 & 34.180 & 3.25 & 8.25 \\
       & 219.261 & 34.115 &         &        &   &   \\
       & 219.211 & 34.246 &         &        &   &   \\
       & 219.228 & 34.180 &         &        &   &   \\

\enddata

\tablenotetext{a} {\it Ulysses \rm /PHEBUS/PVO triangulation error box}
\tablenotetext{b} {\it Ulysses \rm /PVO/WATCH triangulation error box}
\tablenotetext{c} {\it Ulysses \rm /BATSE and \it Ulysses \rm /PVO annuli intersect
at grazing incidence; error box is defined by the four intersections of the \it Ulysses
\rm/BATSE annulus and the WATCH error circle}
\tablenotetext{d} {\it Ulysses \rm /BATSE and \it Ulysses \rm /PVO annuli intersect
at grazing incidence; the annuli in turn graze the WATCH error circle.  The error
box is defined by the three intersections of the \it Ulysses
\rm/BATSE annulus and the WATCH error circle}
\tablenotetext{e} {\it Ulysses \rm /PVO/WATCH triangulation error box}

\end{deluxetable}

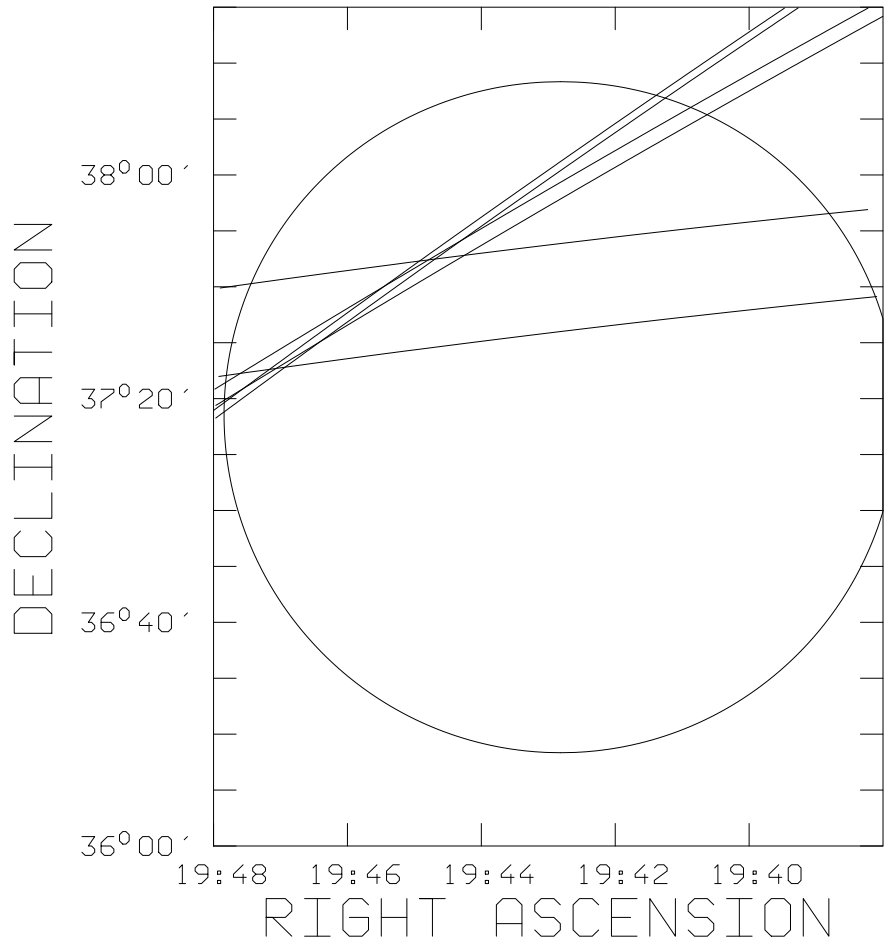
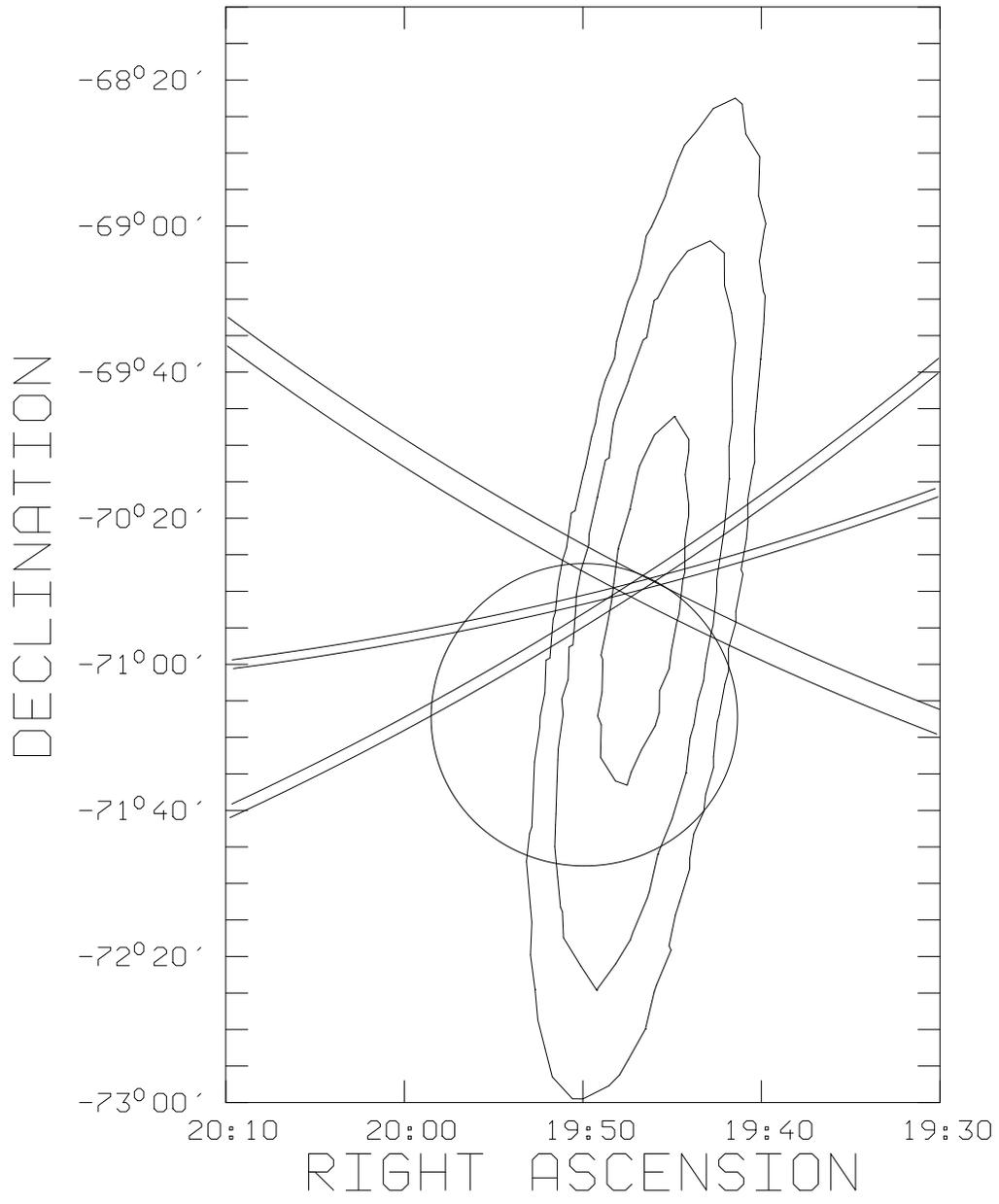
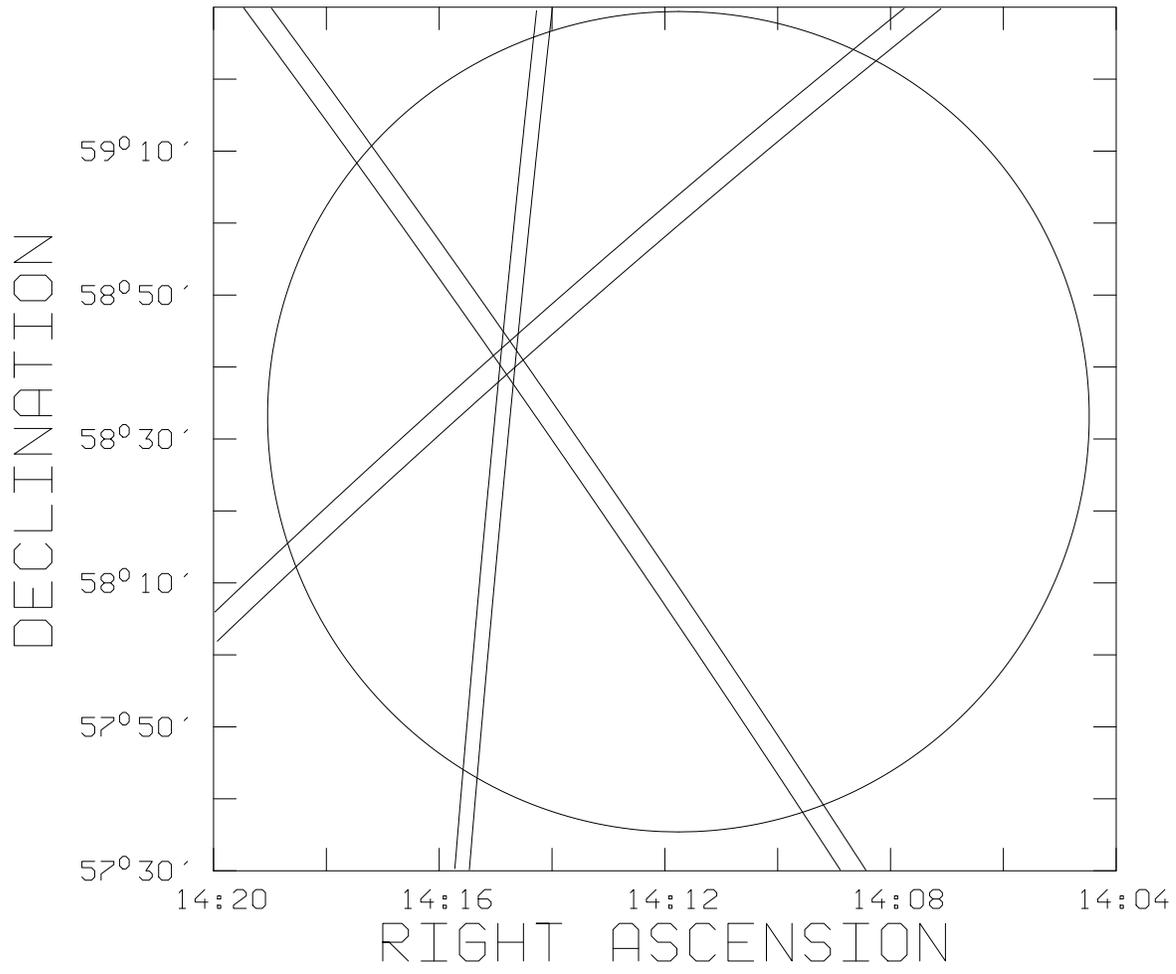
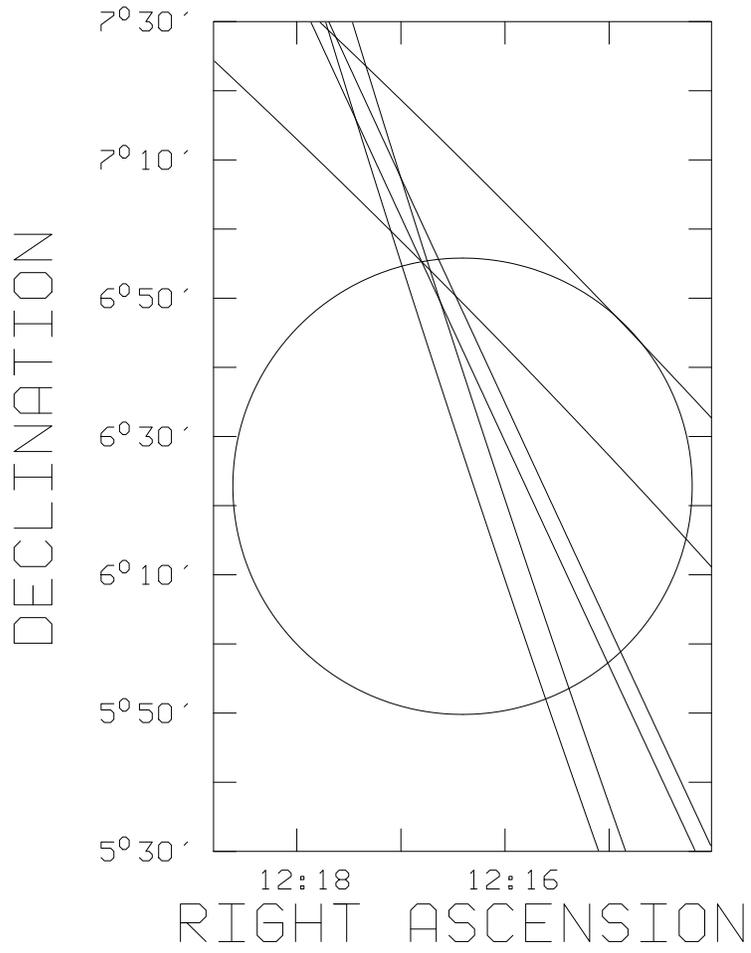
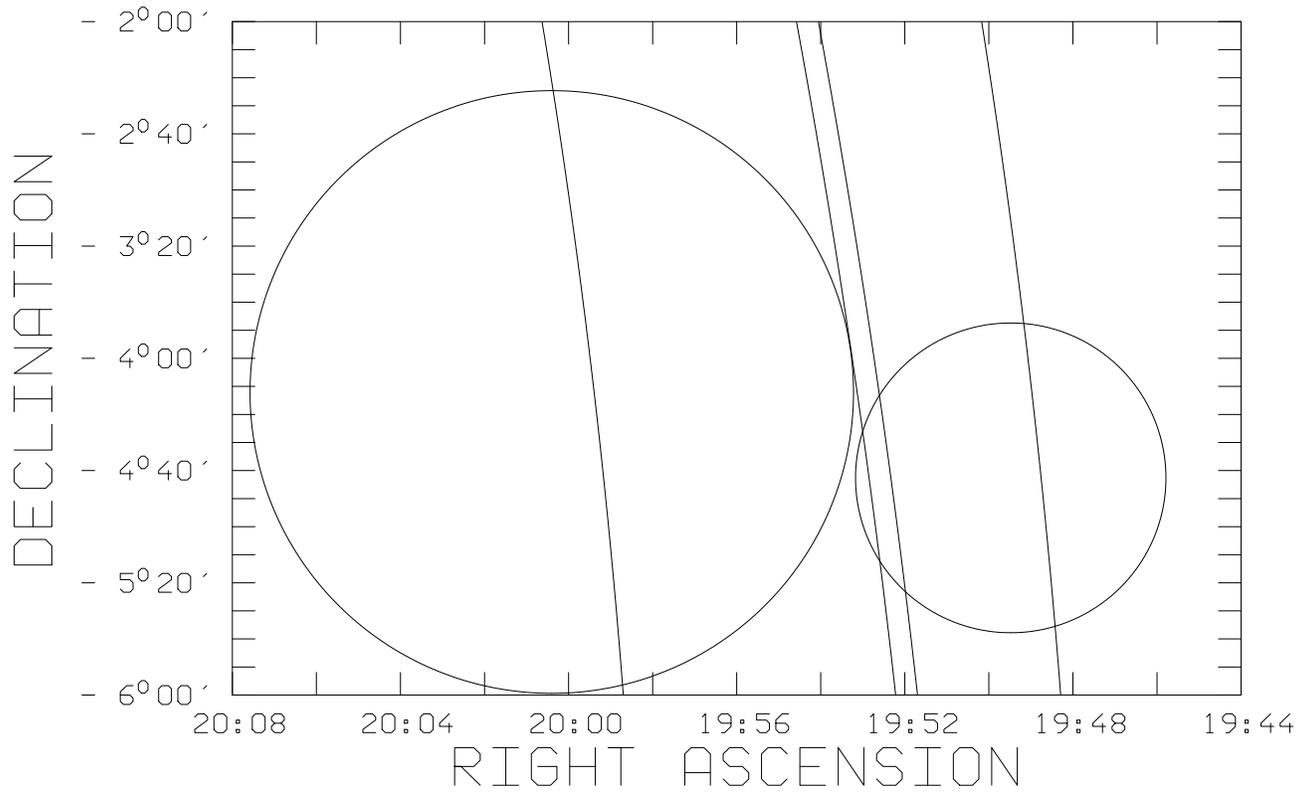
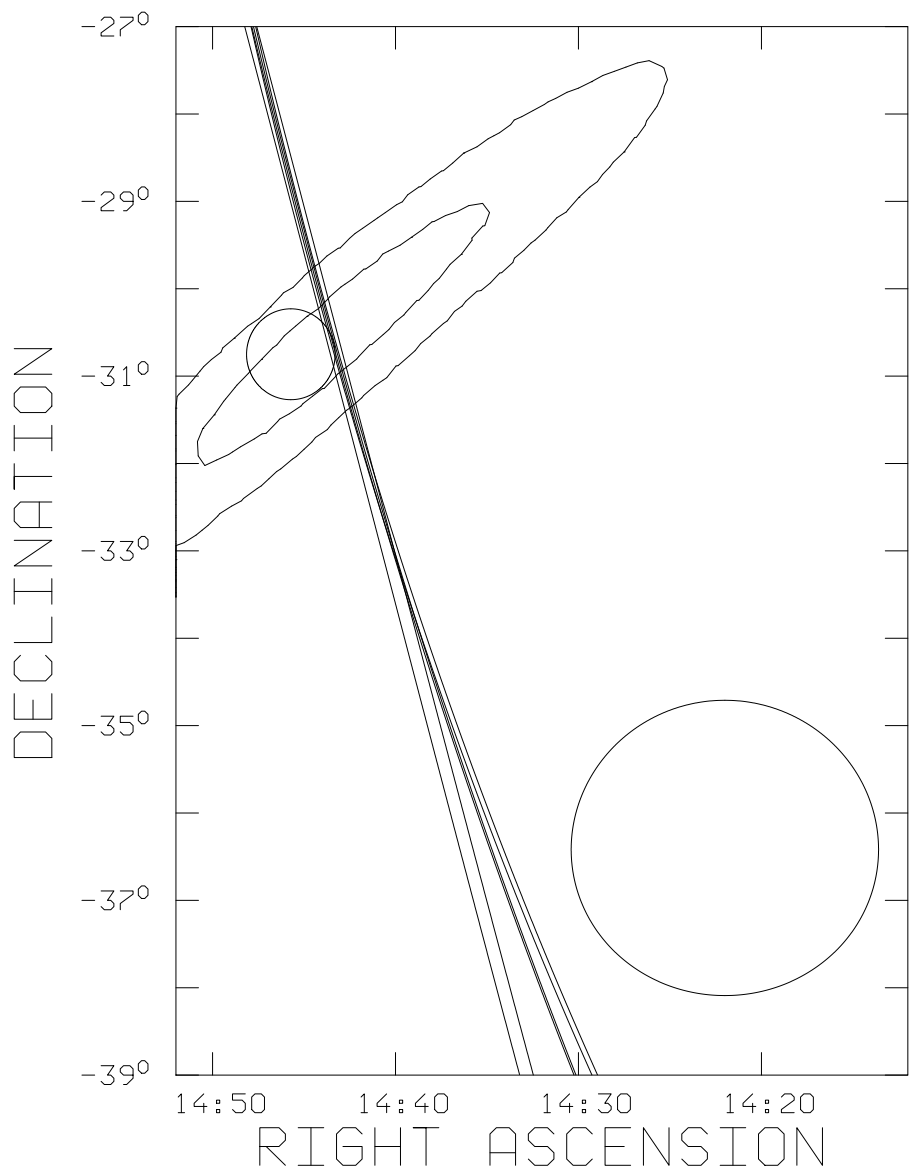
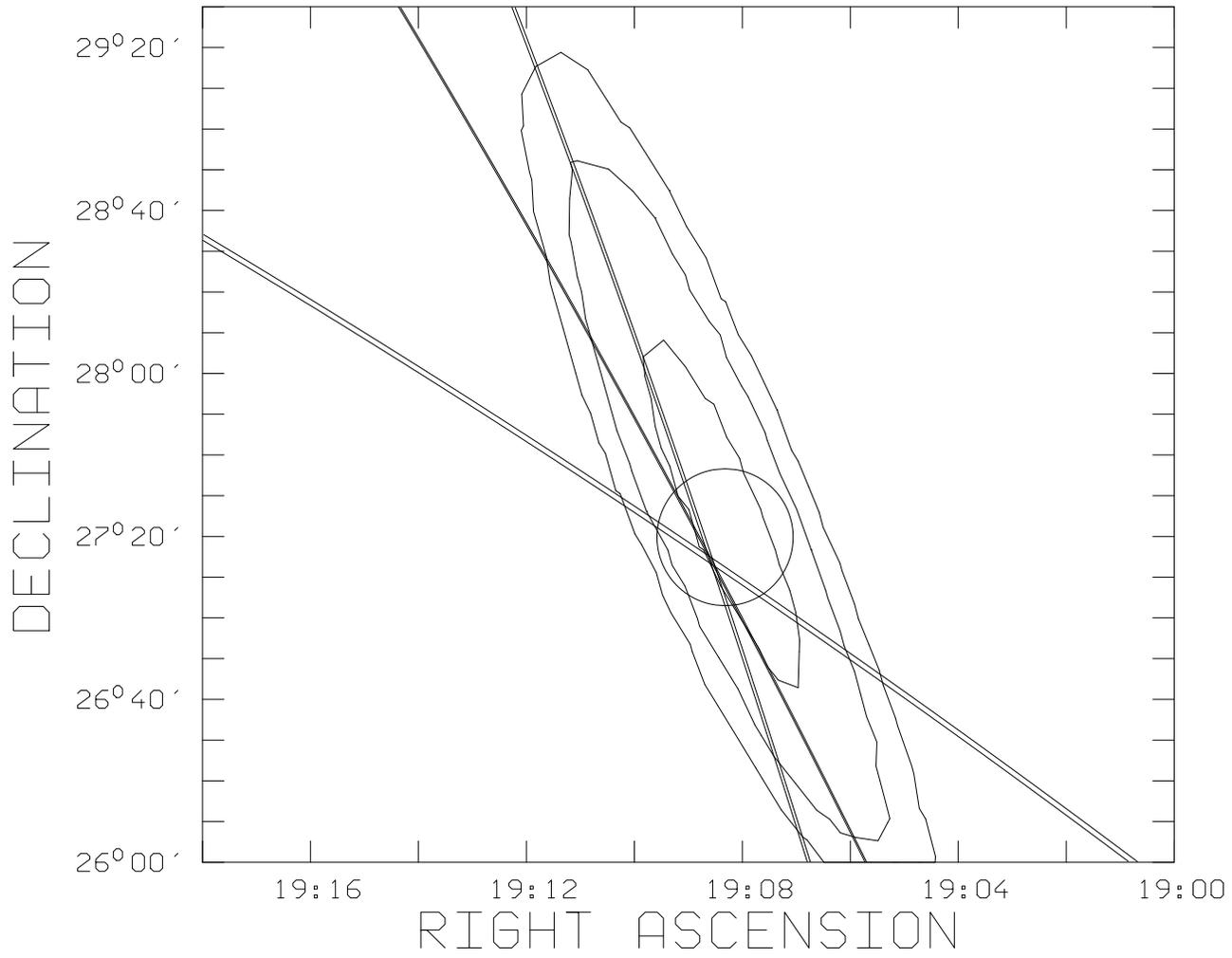
\end{document}